# A high-resolution prediction dataset for solar energy across China (2015–2060)

Daoming Zhu[1], Xinghong Cheng[1*], Yanbo Shen[2*], Chunsong Lu[3], Duanyang Liu[4], Shuqi Yan[4], Naifu Shao[5], Zhongfeng Xu[6], Jida Peng[7], Bing Chen[8]

[1]State Key Laboratory of Severe Weather Meteorological Science and Technology (LaSW), Chinese Academy of Meteorological Sciences, Beijing 100081, China.
[2]China Meteorological Administration Public Meteorological Service Center, Beijing 100081, China.
[3]Nanjing University of Information Science and Technology, Nanjing 210044, China.
[4]Nanjing Innovation Institute for Atmospheric Sciences, Chinese Academy of Meteorological Sciences-Jiangsu Meteorological Service, Nanjing 210041, China.
[5]Tianjin Climate Center, Tianjin 300074, China.
[6]Institute of Atmospheric Physics, Chinese Academy of Sciences, Beijing 100029, China.
[7]Fujian Institute of Meteorology Sciences, Fuzhou 350001, China.
[8]Institute of Plateau Meteorology, CMA, Chengdu 610213, China.

Correspondence to: Xinghong Cheng (cxingh@cma.gov.cn) and Yanbo Shen (shenyb@cma.gov.cn)

**Abstract**:

A high spatiotemporal resolution and accurate middle-to-long-term prediction data is essential to support China's dual-carbon targets under global warming scenarios. In this study, we simulated hourly solar radiation at a 10 km × 10 km resolution in January, April, July, and October at five-year intervals from 2015 to 2060 across China using the WRF-Chem model driven by bias-corrected CMIP datasets and future emission inventories. We further calculated the monthly photovoltaic power potentials based on an improved assessment model. Results indicate that the WRF-Chem model can reproduce the spatiotemporal evolution of solar radiation with small simulation errors. GHI in 2030 and 2060 over China are characterized by a pronounced west-to-east gradient. The interannual fluctuations of GHI from 2015 to 2060 over China's major PV power generation bases are small, and the interannual variability of GHI is mainly dominated by TCC and the influence of

AOD is limited. National averaged PV power generation in China shows a significant growth trend and increases from 68.7 TWh in 2015 to 129.7 TWh in 2060, which is approximately twice the 2015 value. The dataset will provide an important scientific basis for renewable energy planning and grid security under China's dual-carbon strategy.

## 1. Introduction

Against the backdrop of global warming, significant changes have occurred in global solar radiation. Since the 1950s, most regions in the world experienced a decline in solar radiation, which reversed into an increase after the 1990s [1–4]. In China, a similar decrease occurred from 1961 to 1990, but this trend did not continue in most areas after 1990 [4–6]. Variations in solar radiation are mainly controlled by clouds and aerosols, which alter the radiative energy balance [7–11]. China is a sensitive area and a significant impact area of global climate change [12]. With continued warming, the future spatiotemporal evolution of solar energy resources will profoundly affect China's energy security, carbon neutrality strategy implementation, and photovoltaic (PV) industry development.

Numerous studies have assessed China's future PV power generation potential. Yang et al. [13] corrected simulation biases of solar radiation using 25 models from the Coupled Model Intercomparison Project Phase 6 (CMIP6) and found that future increases will be more pronounced under low-emission scenarios, especially in North and Southeast China, which face a high electricity demand despite having relatively low solar radiation. Lei et al. [14] utilized four climate model datasets from the international CovidMIP project and applied a multivariate collaborative bias correction method to assess the impacts of future climate change on solar and wind energy potential under deep decarbonization and carbon neutrality scenarios. Based on 16 CMIP6 models and multivariate bias correction algorithms, Chen et al. [15] found that PV power generation volatility will decline in low-latitude regions, indicating more stable generation, whereas high-latitude regions exhibited significantly increased instability, making them more susceptible to climate change. Chen et al. [16] evaluated 14 models from NASA's Global Daily Downscaling Project climate model ensemble and found that under two Shared Socioeconomic Pathway (SSP2-4.5 and SSP5-8.5), the annual average PV potential decreased in western China but increased in the southeast. Guo et al. [17] combined CMIP6 model outputs with the physical PV model under different SSP scenarios, predicting that annual PV generation could decrease by up to 4% in northern China and the Qinghai–Tibet Plateau under the high-emission scenario (SSP5-8.5), while southern and central regions would show significant growth. Using the WRF-Chem model to simulate future (2046–2050) and history (2016–2020) solar radiation conditions, Zhang et al. [18] demonstrated that the model accurately reproduces China's solar radiation patterns. And a nationwide PV potential increase by 1–4%, attributing primarily to emission reductions, which exceed the negative effect of meteorological conditions.

However, previous studies on estimating the future PV power generation potential in China still exhibit several limitations. First, most studies use the low-resolution climate prediction products and make it difficult to meet the needs of renewable energy development planning and power prediction for PV plants. Second, most studies used the corrected or uncorrected integrated CMIP6 prediction data other than the downscaling coupled meteorology-chemistry models to predict the refined and accurate solar radiation in the future. Furthermore, few studies account for the direct and indirect aerosols radiative effects due to the spatial and temporal evolution of future emission sources, which further affect the predicting accuracy of solar radiation and PV power potential. Finally, many research neglect the influence of the exploitable area in different regions of China, the shading effect between PV panels in PV plants, the optimal inclination angle, and the evolution of the photoelectric conversion efficiency (PEC) and system performance coefficient (SPC)—all of which are critical for accurate estimation of PV potential.

To address the above limitations, we used the online coupled meteorology-chemistry-aerosol model一WRF-Chem driven by bias-corrected CMIP datasets and future emission inventories, and an improved assessment model of PV power potentials to develop a high spatiotemporal resolution and accurate middle-to-long-term solar energy prediction dataset. The dataset accounts for cloud–aerosol–radiation interactions, and including the evolution of future emission sources, PEC and SPC, as well as the combined effects of developable area and PV shading. It provides essential baseline data and scientific support for the scientific formulation and implementation of China's renewable energy development plans, which aiming at achieving the dual carbon goals, mitigating climate risks in the PV industry, and ensuring grid stability and security. On one hand, solar resources exhibit pronounced spatial heterogeneity [19]. High-resolution data can more accurately capture the availability of solar resources across regions, offering a robust basis for siting distributed solar projects. On the other hand, solar resources also demonstrate marked seasonal and interannual variability [20]. Long-term prediction data can effectively predict future solar resource trends, providing forward-looking guidance for energy policy formulation and industrial layout optimization. This supports a dynamic balance between electricity supply and demand while enhancing the grid resilience to extreme climate events. Finally, the incorporation of cloud-aerosol-radiation feedback mechanisms allows the dataset to simulate future climate scenarios with greater realism, thereby offering crucial scientific support for achieving the national carbon peaking and carbon neutrality targets.

## 2. Methodology and data

### 2.1. Model setup

We made a long-term regional simulation by using a fully meteorology-chemistry coupled model, WRF-Chem (the Weather Research and Forecasting model coupled with Chemistry) to construct a high-resolution prediction dataset for solar radiation and PV potential over China in January, April, July, and October at five-year intervals from 2015 to 2060. The simulations were driven by the monthly emission inventory at 5-years intervals for the carbon neutral scenario with a spatial resolution of 0.25°×0.25° from the Dynamic Projection model for Emissions in China (DPEC) [21], and the 6-hourly multi-model integrated and bias-corrected GCM dataset using the ensemble mean of the 18 CMIP6 models with a resolution of 1.25 degree [22]. The bias correction of GCM datasets were performed using the ERA5 reanalysis dataset, and the projections incorporated the ensemble mean of 18 models from the CMIP6. The DPEC model simulated future technological evolution processes for various pollution sources and their impact on emissions under different socioeconomic development and policy evolution scenarios [23]. This study adopts the carbon-neutral climate scenario from the second scenario dataset (v1.1) of the DPEC model, spanning 2015 to 2060, to better simulate China's future climate in the context of carbon peaking and carbon neutrality. Ultimately, the WRF-Chem model is used for downscaling simulations, comprehensively considering cloud-aerosol-radiation interactions to enhance the dataset's resolution and accuracy. Figure 1 shows the specific technical roadmap for simulation of solar radiation and calculation of PV power potentials.

The simulation domain (as shown in Figure 2) covers the whole China and its surrounding areas using the double-nested simulation grids with horizontal resolutions of 50 km and 10 km, and with 41 vertical layers of varying thickness (between the surface and 10 hPa). Initial and boundary conditions for the meteorological elements were from the above multi-model integrated and bias-corrected GCM dataset, and those for the chemical species in the WRF-Chem were extracted from the output of the MOZART-4 global chemical transport model [24]. The simulation for every day was conducted from 0000 UTC on previous day to 0000 UTC on next day, with a 24-h spin-up time. Outputs from the WRF-Chem model at 0000 UTC on current day were used as the initial and boundary conditions for the entire run. The following parameterization schemes of physical processes within the WRF-Chem model were used in this study: the improved Morrison two-moment microphysics scheme [25–27], the RRTMG longwave and shortwave radiation scheme [28], the revised Monin Obukhov surface layer scheme [29], the Grell 3D ensemble convective parameterization scheme [30], the RUC land-surface scheme [31], and the Shin-Hong atmospheric boundary layer scheme [32].

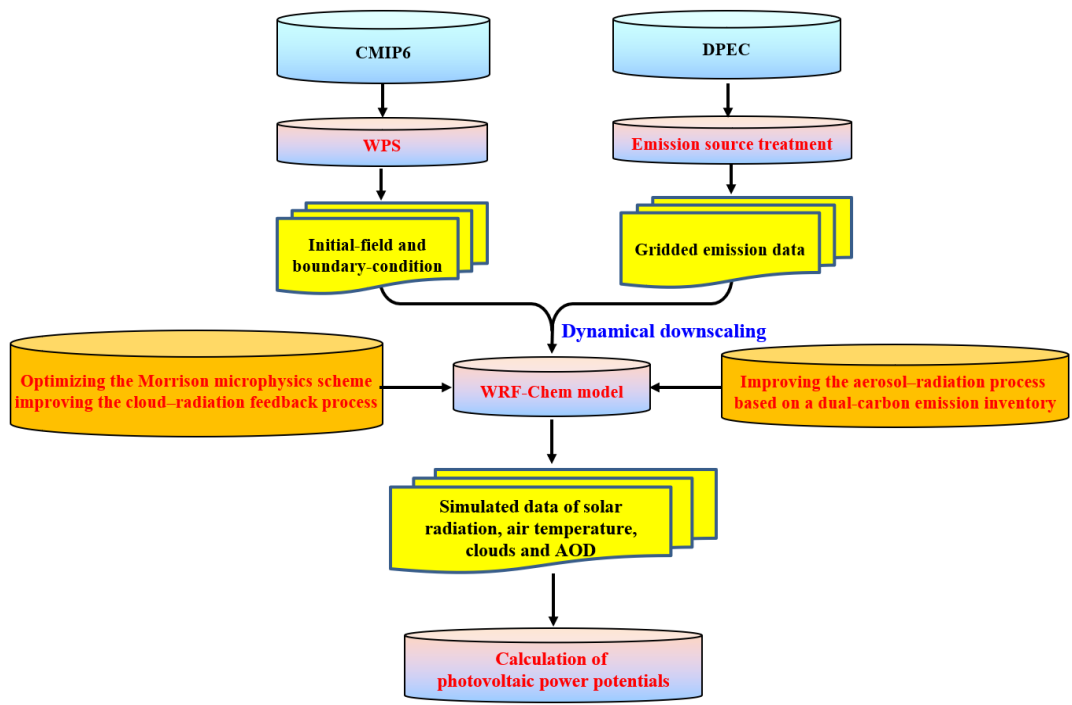

**Fig. 1** Technology roadmap of middle-to-long-term prediction method for solar radiation and PV power potentials

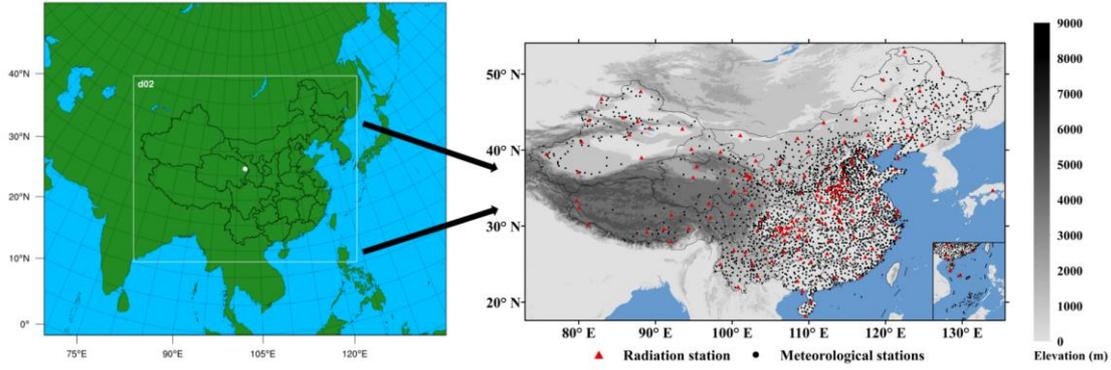

**Fig. 2** Double-nested model domains (a) and distribution of in-situ radiation and meteorological stations (b) within the inner domain (marked by D02), which covers China and its surroundings.

**2.2 Assessment model of PV power**

We improved the assessment model of PV power generation potential based on the method suggested by Li and He et al. [13,33]. Our model includes the future evolution of PCE and SPC, and the impacts of the exploitable area and the shading of PV arrays on evaluation of PV power. Improvements include that 1) undevelopable land is excluded when calculating effective PV area to improve spatial accuracy referring to the results from Wang et al. [34]; 2) following the approach of Fernández-Infantes et al. [35–37], the model incorporates shading effects between adjacent solar panels; 3) the impact of future technological advancements is represented through the average PCE and SPC at five-year intervals from 2015 to 2060 (Table 1) suggested by Lu et al. [38].

The calculation formula of PPP is written as follows:

$$P = S \times [\eta_0 \times \eta_{loss} \times (1 - \alpha_p)] \times A_\rho \times A_d \quad (2)$$

where $S$ represents China's annual average GHI; $\eta_0$ denotes the average PCE; $\eta_{loss}$ is SPC which accounts for efficiency losses due to dust, snow, and haze shading, as well as inverter, transformer, and circuit losses; $\alpha_p$ is the panel albedo, assumed to be 0.1 [33]; $A_\rho$ denotes the developable land area suitable for PV power generation; and $A_d$ represents the shading effect coefficient between solar panels.

Table 1 The PEC and SPC at five-year intervals from 2015 to 2060 in China.

| Year | PCE (%) | SPC (%) |
|---|---|---|
| 2015 | 16.0 | 81.6 |
| 2020 | 19.4 | 83.1 |
| 2025 | 20.9 | 84.6 |
| 2030 | 22.4 | 86.1 |
| 2035 | 23.4 | 87.1 |
| 2040 | 24.4 | 88.1 |
| 2045 | 25.4 | 89.1 |
| 2050 | 25.9 | 89.6 |
| 2055 | 26.4 | 90.1 |
| 2060 | 26.9 | 90.6 |

**2.3 Stability Calculation**

The stability of solar radiation is a key factor in evaluating the PPP. We used the calculation method of Coefficient of Variation ($CoV$) [39–41] to assess the stability of GHI. The $CoV$ is defined as the ratio of the standard deviation to the mean, and is widely applied to evaluate the variability of solar energy resources across spatial and temporal dimensions. The formula is written as follows:

$$CoV = \frac{1}{\bar{X}} \sqrt{\frac{1}{N} \sum_{i=1}^{N} (X_i - \bar{X})^2} \times 100 \qquad (1)$$

where $X_i$ and $\bar{X}$ is the $i$-th GHI and mean of GHI over a period, respectively. N is the total number of samples.

We analyzed the intra-annual and interannual stability of GHI using hourly GHI data. The intra-annual $CoV$ represents the fluctuation in solar radiation within a single year, which is primarily driven by Earth's revolution and seasonal variations [42,43]. The interannual $CoV$ reflects variations in solar radiation induced by climate anomalies and emission changes [20]. When calculating interannual $CoV$ values, $X_i$ represents the annual average GHI which is calculated with the data in four months and N denotes the simulation interval used in this study (10 years).

## 3. Results

### 3.1 Validation of simulation results

We evaluated the accuracy and reliability of solar radiation and its key determinants such as clouds and aerosols simulated by the WRF-Chem model using measured GHI, Ta, TCC data at radiation and surface meteorological stations, as well as AOD products retrieved from the MODIS satellite for January, April, July and October of 2015 and 2020. Evaluation indexes include correlation coefficient (R), root mean square error (RMSE), mean bias (MB), normalized mean bias (NMB), and index of agreement (IOA).

Figure 3(a-e) shows scatter plots between the monthly average of GHI, DNI, DIF, Ta, TCC, AOD simulated by the WRF-Chem model and observations for the four representative months of 2015 and 2020. The correlation coefficients between simulated and observed GHI, DNI, DIF, Ta, TCC, and AOD all exceed 0.61 and are statistically significant at the 99.9% confidence level. IOA values for most elements are above 0.75 except for AOD, indicating that the modeled spatiotemporal variations of surface radiation and temperature, and cloud cover are consistent with observations. IOA for AOD is relatively smaller and it may be related with the retrieved errors of satellite data and simulated bias caused by the uncertainty of emission inventory and climate fields. In general, the model overestimate GHI, DNI, DIF, Ta, and TCC, particularly GHI, DNI and TCC, whereas it significantly underestimates the AOD.

The greater dispersion between simulation and observations may be attributed to uncertainties in macro- and microphysical properties of cloud in the CMIP6 climate fields and the DPEC emission inventory, and the complexity of cloud–aerosol–radiation interactions [22,44]. Overall, the WRF-Chem model can reproduce spatiotemporal variations of solar radiation and its key determinants, providing a reliable foundation for subsequent analysis of PPP.

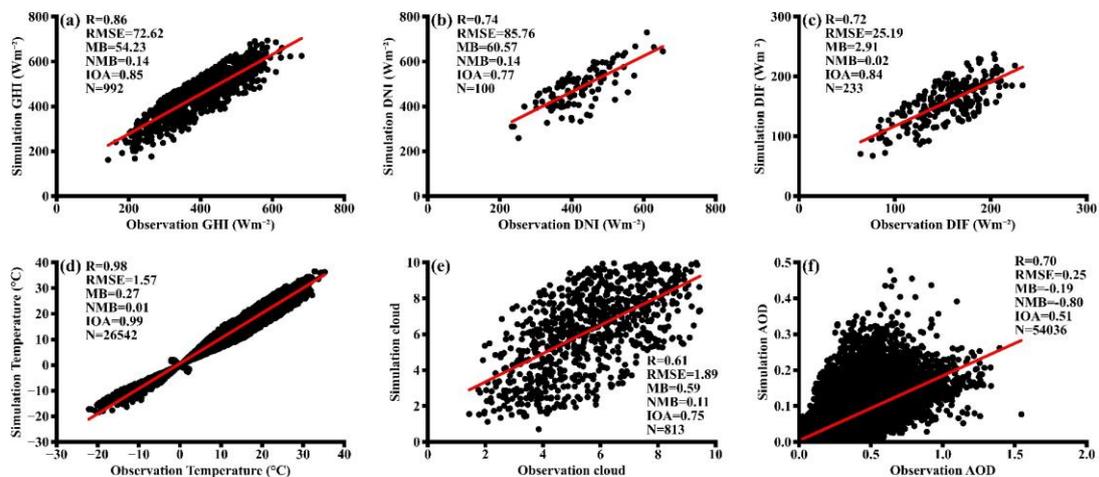

**Fig. 3** Scatter plots of simulated GHI, DNI, DIF, Ta, TCC, and AOD and observations in January, April, July, and October of 2015 and 2020 over China.

**3.2 Spatial distribution of solar radiation**

Figure 4 illustrates the spatial distribution of annual averaged GHI and DIR in 2030 and 2060 over China, along with their changes relative to 2020. Results show that high GHI zones are primarily concentrated in Tibet, Qinghai, Xinjiang, and Gansu provinces, with GHI reaching up to 580 W·m$^{-2}$. Whereas, low-value zones are mainly located in the Northeast China, with GHI around 210 W·m$^{-2}$.

High-value areas for DIR are predominantly distributed across Xinjiang, Inner Mongolia, Gansu, and Tibet, where DIR reaches approximately 450 W·m$^{-2}$ due to high altitude and less cloud cover. Low-value areas are located in Northeast China, Sichuan and Guizhou provinces, where basin topography and frequent cloud cover substantially reduce DIR, resulting in DIR being as low as 40 W·m$^{-2}$. Relative to 2020, both GHI and DIR show a declining trend in Southwest and Northeast China by 2030, while Xinjiang, the Beijing-Tianjin-Hebei (BTH) region, and the southeastern coastal areas show significant increases. By 2060, radiation values in the southwest, northeast, and southeastern coastal regions decline markedly compared to 2020, whereas GHI and DIR in the BTH region increase remarkably. This spatial pattern may be related with expected decrease in emission sources, underscoring the substantial influence of aerosol changes on solar radiation distribution across China.

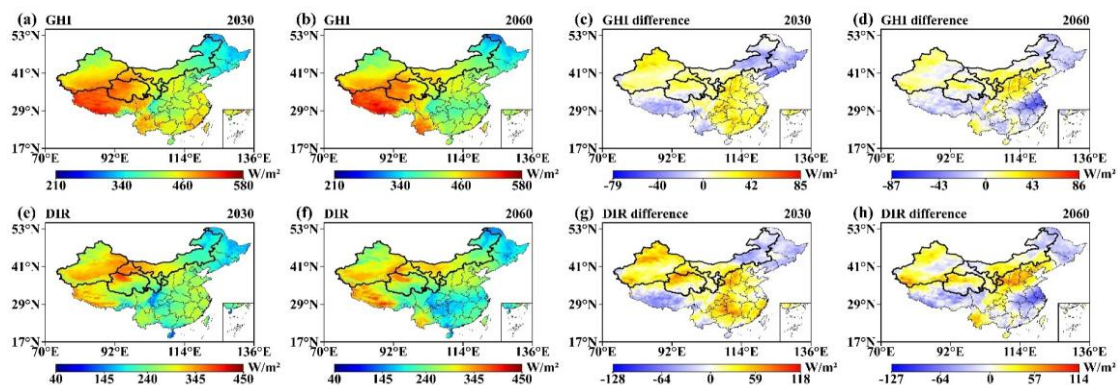

**Fig. 4** Spatial distribution of annual averaged GHI and DIR in 2030 and 2060 over China, along with their changes relative to 2020. Differences are calculated as values in 2030 (or 2060) minus those in 2020.

Figure 5 presents the spatial distribution of monthly averaged GHI in January, April, July, and October of 2030 and 2060 over China, as well as their changes relative to 2020. In January 2030, high GHI zones are primarily concentrated in Tibet, Qinghai, and the southeastern coastal regions, while low zones are found in Xinjiang, Inner Mongolia, and Northeast China, with minimum values around 75 W·m$^{-2}$. In April, GHI is the highest in western China and the lowest in northeast China. By July, high GHI areas shift toward the northwest and southeast, with peak values reaching approximately 715 W·m$^{-2}$, while GHI in northeast China remains the lowest values. In October, high GHI values are centered in Tibet, while those in northeast China retains low values. By 2060, the overall spatial distribution of GHI remains broadly similar to 2030. However, GHI in July over southeast China shows a significant decline compared with 2020.

  Relative to 2020, changes in monthly averaged GHI across China in 2030 and 2060 show a distinct seasonal and regional variation characteristic, indicating that PV power generation will face with huge challenge in the context of climate change. In January, GHI decrease in the Tibet, southwest and central China, whereas those increase in northeast China, BTH regions, northern Xinjiang and the eastern coastal areas. By April, GHI decrease in northeast China but generally increase in the northwest and central regions. In July, GHI decrease in the northeast and southwest areas while increasing along the southeastern coast and in BTH region. By October, GHI decline in the southern and northeastern regions, with those in BTH area showing an upward trend. Overall, changes in GHI across months and regions are likely influenced by multiple interacting factors, including climate change, evolving emission sources, and variations in surface vegetation cover, and so on.

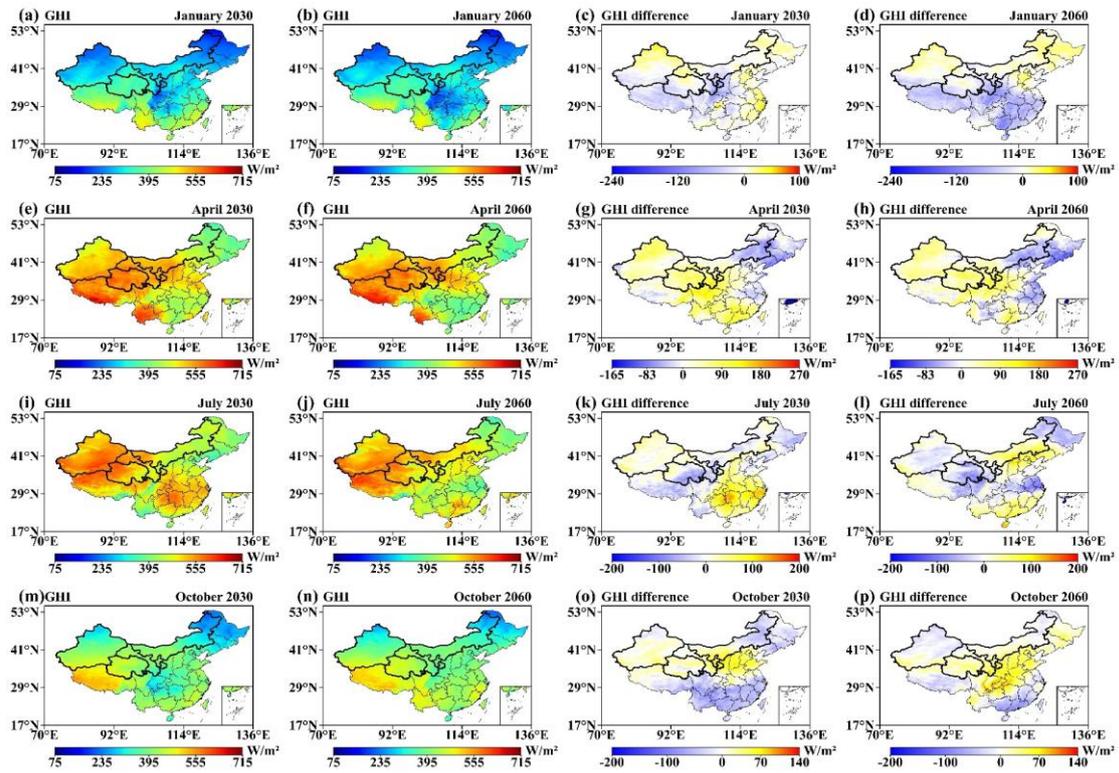

**Fig. 5** Spatial distribution of monthly averaged GHI in January, April, July, and October of 2030 and 2060 over China, along with their changes relative to the corresponding month in 2020. Differences are calculated as values in 2030 (or 2060) minus those in 2020.

Figure 6 presents the spatial distribution of monthly averaged DIR in January, April, July, and October of 2030 and 2060 across China, along with their variations relative to the same month in 2020. Similar to GHI, the spatial patterns of DIR exhibit clear seasonal and regional variation characteristics. In January 2030, high DIR zones are primarily concentrated in Tibet, Yunnan, and the southeastern coastal areas, while low-value zones are distributed across Xinjiang, Sichuan, and the Northeast. By April, high-value zones shift northwestward, concentrating in western China, while low-value zones move toward the southeastern coast. By July, the highest values occur in the northwest and along the southeast coast, with peak values reaching 640 W/m², while DIR in the northeast maintains the lowest values. In October, high-value zones are centered in Tibet and southern Xinjiang, whereas low-value zones are distributed across Xinjiang, Sichuan, and the northeast. Except for July, spatial distribution patterns of DIR in other months of 2060 are similar to those in 2030. DIR in July of 2060 over southeastern China experiences a significant decline compared to those in 2030.

Compared with 2020, changes of monthly averaged DIR in 2030 also exhibit pronounced spatial heterogeneity. In January, DIR in most regions declines particularly in the southwest and central parts of China, whereas those in northern Xinjiang and the eastern coastal areas increase. By April, decreases of DIR are concentrated in the northeast. In July, DIR decreases in the northeast and southwest but rise substantially in the southeastern coastal regions and the BTH area. By October, DIR decline across southern and northeastern China, while those increase in the BTH region.

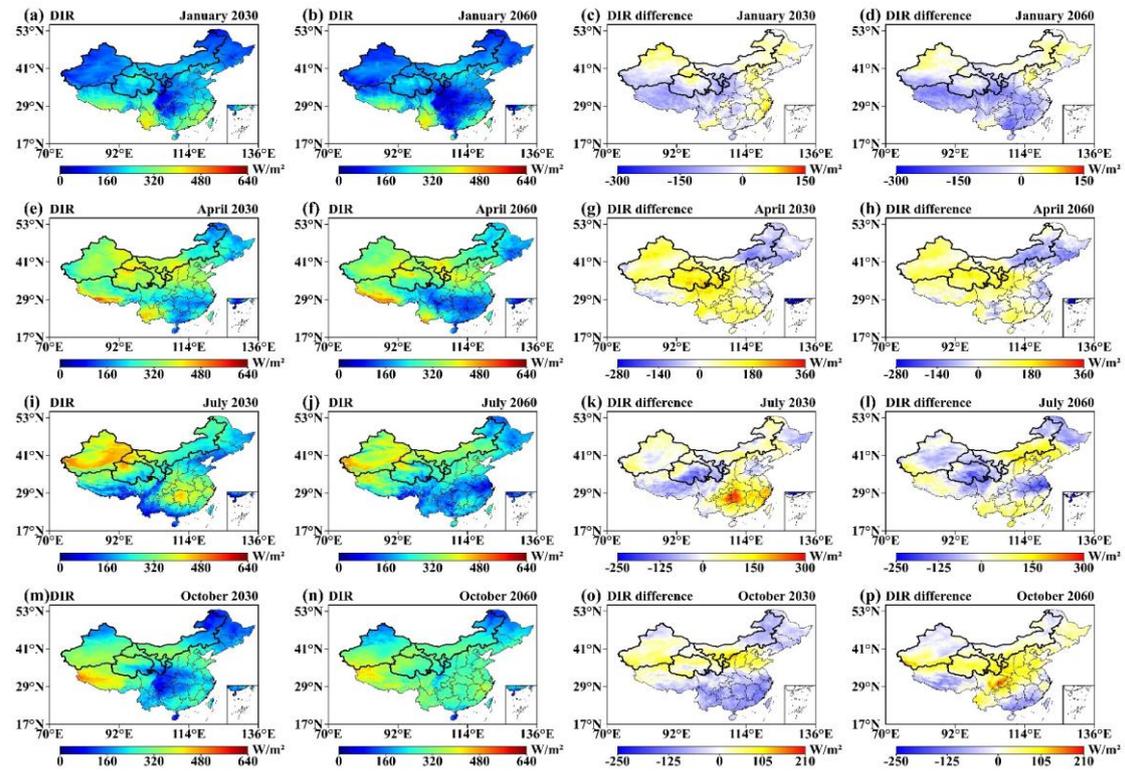

**Fig. 6** Same to Figure 5, but for DIR.

### 3.3 Temporal variation of solar radiation

Figure 7 illustrates the interannual variation of annual and monthly averaged GHI and DIR from 2015 to 2060. As shown in Figure 7(a), both GHI and DIR exhibit similar fluctuations, characterized by an overall upward trend with two notable peaks around 2020 and 2030, and two troughs in 2025 and 2045. The maximum of GHI and DIR occurs in 2030, reaching the minimum around 2045. Figures 7(b) and 7(c) further depict the interannual variability of GHI and DIR in different seasons. The GHI exhibits relatively stable variations in the same month, and those in April and July are significantly higher than October and January. In contrast, DIR demonstrates more pronounced interannual variations for the same month, and those in April, July, and October are obviously higher than January. Overall, interannual fluctuations are smaller for GHI but slightly larger for DIR, reflecting the stronger sensitivity of DIR to atmospheric transparency and scattering conditions.

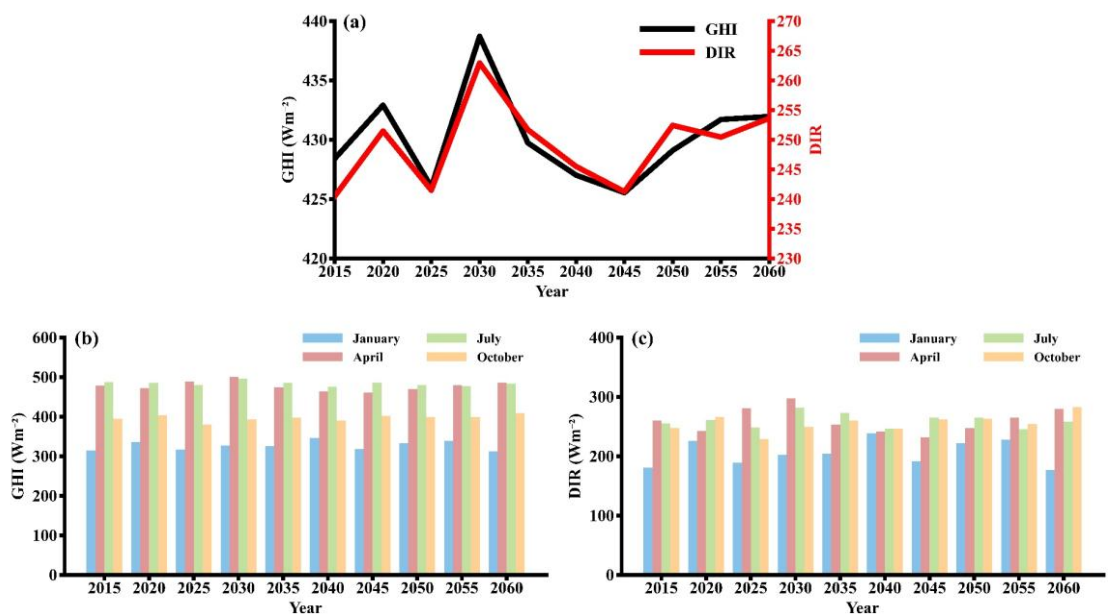

**Fig. 7** Interannual variations of annual average (a) and monthly mean (b, c) of GHI and DIR over China from 2015 to 2060.

To further interpret the interannual fluctuations of annual averaged GHI, we give the interannual variation characteristics of national annual averaged GHI and TCC, AOD from 2015 to 2060 in China (Figure 8). A strong negative correlation between GHI and TCC is observed. An increase in TCC is accompanied by a decrease in GHI, reflecting the significant reduction of surface solar radiation by cloud cover through reflection and scattering [45]. This attenuation effect is particularly evident in these years with extensive cloud cover.

The AOD gradually decreases following the control of emission sources over the simulation period. GHI exhibits a negative correlation with AOD while the correlation is not significant compared with that between GHI and TCC. This indicates that AOD has a weaker moderating effect on GHI than TCC, likely due to the complex direct and indirect aerosol radiation effects.

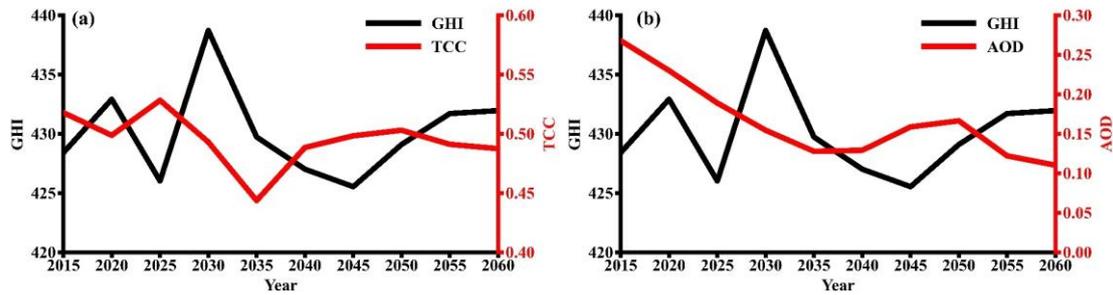

**Fig. 8** Time series of national annual averaged GHI and TCC(a), AOD(b) from 2015 to 2060 in China.

**3.4 Stability of solar radiation**

Stability is crucial for the reliability and performance of PV power generation. We analyzed the spatial distribution characteristics of intra-annual stability in GHI across China for 2020, 2030 and 2060 (Figure 9). $CoV$ in three years consistently show a clear latitudinal dependency: low-latitude regions generally have higher intra-annual stability and high-latitude regions experience larger variations in solar radiation. And it is mainly due to the regional difference of seasonal fluctuations in GHI. Additionally, GHI in Sichuan, Chongqing, and Guizhou show low stability and it may be related with their complex topography and frequent cloud cover. Persistent clouds enhance the reflection and scattering of solar radiation, increasing its variability and resulting in higher coefficients of variation ($CoV$). Consequently, these areas exhibit poorer intra-annual stability of GHI. Such spatial difference in intra-annual stability provides valuable guidance for the planning and deployment of PV power stations. In low-stability areas, it is essential to incorporate additional energy storage systems or backup power sources to maintain the continuity and reliability of electricity supply.

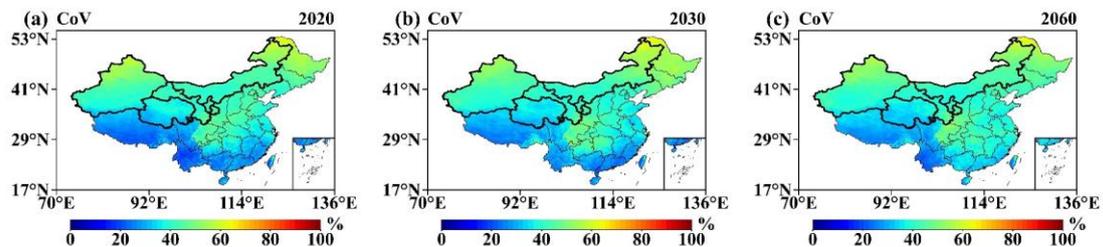

**Fig. 9** Spatial distribution of intra-annual stability of GHI in 2020, 2030 and 2060 over China.

Figure 10 gives the spatial distribution of interannual stability in GHI from 2015 to 2060 across

China. Northeast, central, eastern, and southern China exhibit significant interannual fluctuations. Variations in cloud cover and surface albedo may be the dominant factor for these fluctuations. As clouds reduce surface radiation through reflection and scattering, and seasonal snow and ice cover lead to high surface reflectivity in the Northeast, which alters the absorption and reflection of solar radiation and further affects the stability of GHI. In contrast, cloud cover in arid regions such as the Tibet Plateau, Inner Mongolia and Xinjiang province is less and results in smaller interannual fluctuations and higher stability in solar radiation, thus provides better conditions for PV power generation. Conversely, GHI in southeastern China with frequent cloudiness shows lower interannual stability, which may affect the reliability of PV electricity output.

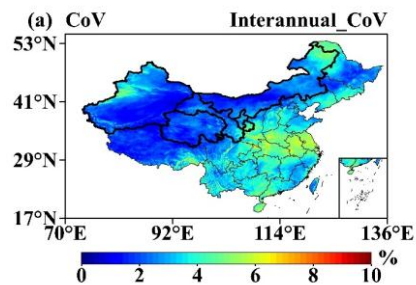

**Fig. 10** Spatial distribution of interannual stability of GHI from 2015 to 2060 over China.

**3.5 Photovoltaic Power Potential**

Figure 11 presents the interannual and seasonal variations of national PV power generation in China from 2015 to 2060. China's PV power generation exhibits a steady upward trend, increasing from 68.68 TWh in 2015 to 129.70 TWh in 2060. The predicted PV power generation in this study is generally lower than previous studies. The estimated power generation in 2020, 2030, and 2060 is close to other researches, while the results for 2015 and 2050 are significantly lower than previous studies. The main reasons for the differences are that previous studies inadequately considered the exploitable area and shading effects, while this study introduced calculation errors by using four representative months for the annual output estimate due to computational limitations. Overall, because this study takes into account the interannual variation of futurePCE and SPC as well as the influence of undevelopable areas and shading effect of photovoltaic array, the estimation results in this study are more reliable.

Figure 11(b) illustrates the seasonal variation characteristics of PV power generation in January, April, July, and October. The results reveal significant seasonal fluctuations, primarily driven by differences in solar zenith angle, sunshine duration, and climatic conditions across seasons. Moreover, monthly PV power generation has shown a notable upward trend, attributable to the continuous advancement of photovoltaic cell technology. PV power generation increased substantially in April and July, significantly higher than that in January and October due to higher solar radiation and longer sunshine hours.

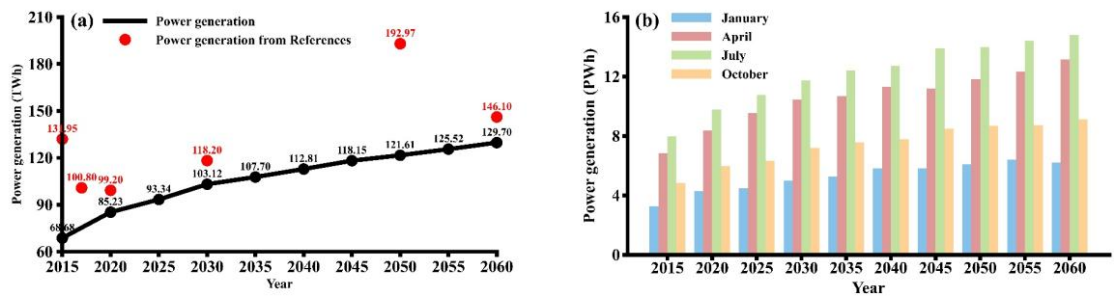

**Fig.11** Time series of annual (a) and monthly (b) PV power generation from this study and other researches at five-year intervals during 2015-2060 in China.

Figure 12 gives the spatial distribution of PV power across China in 2020, 2030, and 2060. Overall, high PV power areas are primarily concentrated in Xinjiang, Tibet, Inner Mongolia, Gansu, Ningxia and the BTH region. These regions have naturally superior conditions for developing PV industry due to their longer sunshine hours and high solar radiation intensity. With continuous advancements in PV technology and developments of electric grid infrastructure, PV power increases markedly from 2030 to 2060, particularly in resource-rich areas such as southern Xinjiang and Xizang, west of Qinghai and Inner Mongolia, most areas of Gansu and Ningxia, where the installed capacity peaks at approximately 2 GW. Growth of PV power from 2020 to 2060 in the above resource-rich areas will be conducive to the development of large-scale PV power base. Although southeastern China also shows improvement in PV power, the overall potential remains limited due to more cloud cover, restricted land availability associated with urbanization.

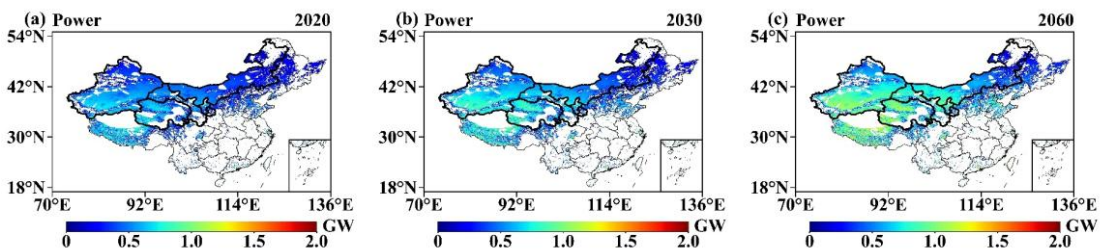

**Fig. 12** Spatial distribution of annual PV power in China in 2020, 2030 and 2060.

Figure 13 illustrates the spatial distribution characteristics of PV power in China during January, April, and July for both 2030 and 2060. In both years, regions with higher PV power are predominantly concentrated in western China, including Xinjiang, the Tibetan Plateau, and Inner Mongolia, and the Beijing-Tianjin-Hebei region. The spatial patterns exhibit distinct seasonal variations: PV power in January and October consistently show relatively low power due to shorter sunshine hours and weaker solar radiation, while those in April and July show substantial increases driven by longer sunshine duration and stronger solar irradiance. With the continuous advancements in PV technology and the upgrading of electric grid infrastructure, PV power in four month of 2060 are obviously higher than those in 2030.

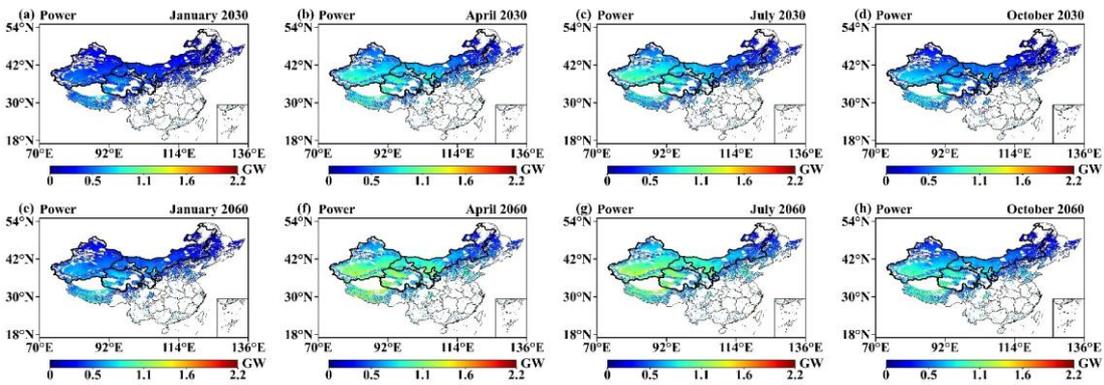

**Fig. 13** Same to Figure 12, but for monthly PV power.

Table 2 summarizes the annual averaged GHI and total PV power for four major PV bases—Beijing-Tianjin-Hebei, Inner Mongolia, Xinjiang, and Qinghai—in 2020, 2030, and 2060. Among these regions, PV power in Xinjiang is the maximum, and its annual PV power increases dramatically from 5,598.39 GW in 2020 to 8,560.67 GW in 2060, underscoring Xinjiang's important role and huge potential in China's PV industry. Inner Mongolia ranks the second, with power rising from 2,925.34 GW to 4,511.08 GW. Although the averaged GHI in Qinghai is the highest among all regions, its total PV power takes third place due to limited developable land area. Meanwhile, the significant decrease in emission sources intensity in the BTH region has led to increased atmospheric transparency, providing favorable conditions for PV power generation. Also, the averaged GHI here increases from 390.06 W/m² in 2020 to 415.64 W/m² in 2060, accompanied by a remarkable rise in PV power. In summary, all four regions show a significant upward trend in both solar radiation and PV generation potential.

**Table 2.**
The annual averaged GHI and total PV power in China's four major PV bases in 2020, 2030 and 2060.

|  | 2020 | | 2030 | | 2060 | |
| --- | --- | --- | --- | --- | --- | --- |
|  | GHI (W/m$^2$) | Power (GW) | GHI (W/m$^2$) | Power (W) | GHI (W/m$^2$) | Power (W) |
| Beijing–Tianjin–Hebei | 390.06 | 440.73 | 400.06 | 540.09 | 415.64 | 710.42 |
| Inner Mongolia | 403.79 | 2925.34 | 398.97 | 3482.71 | 408.34 | 4511.08 |
| Xinjiang | 438.28 | 5598.39 | 453.17 | 6890.14 | 444.52 | 8560.67 |
| Qinghai | 475.10 | 2109.92 | 482.47 | 2582.59 | 477.18 | 3229.75 |

## 5. Summary and discussions

We used the WRF-Chem model, the multi-model integrated and bias-corrected GCM dataset from CMIP6 and the future emission inventory from DPEC to accurately simulate GHI, DNI, DIF, Ta, TCC and AOD at high spatiotemporal resolution at five-year intervals from 2015 to 2060 across China. We also constructed a more precise estimation model for PPP, explicitly accounting for future variations in PCE and SPC, land availability for PV development, and shading effects within PV arrays. We assessed the spatiotemporal evolution of future PPP and developed a high-resolution solar energy prediction dataset in China covering 2015–2060.

The WRF-Chem model can reproduce spatiotemporal variations of solar radiation and its key determinants, providing a reliable foundation for calculations of PV power potentials. High GHI zones are primarily concentrated in Tibet, Qinghai, Xinjiang, and Gansu provinces, while low-value zones are mainly located in Northeast Chin in 2030 and 2060. Compared to 2020, GHI and DIR in 2030 and 2060 over the southwest and northeast China decline, whereas those in Xinjiang and BTH regions increase remarkably. Changes in monthly averaged GHI and DIR in 2030 and 2060 show a distinct seasonal and regional variation characteristic, indicating that PV power generation will face a huge challenge in the context of climate change. The national annual average of GHI and DIR in China show a fluctuating upward trend from 2015 to 2060, with two notable peaks around 2020 and 2030, and two troughs in 2025 and 2045. Interannual fluctuations of GHI in four months are smaller but slightly larger for DIR, and the interannual variability of GHI is mainly dominated by TCC and the influence of AOD is limited. Intra-annual stability in GHI across China for 2020, 2030 and 2060 consistently show a clear latitudinal dependency. GHI in Northeast, central, eastern, and southern China exhibit significant interannual fluctuations, while interannual variations of GHI in the Tibet

Plateau, Inner Mongolia and Xinjiang provinces are smaller. China's PV power generation exhibits a steady upward trend, increasing from 68.68 TWh in 2015 to 129.70 TWh in 2060. The predicted PV power generation in this study is generally lower than previous studies due to inadequate consideration of the exploitable area and shading effects in other researches, and the introduced calculation errors by using four representative months to estimate the annual total PV output in this study. Annual and seasonal total PV power both increase markedly from 2030 to 2060, particularly in Xinjiang and Xizang, west of Qinghai and Inner Mongolia, most areas of Gansu and Ningxia, and the BTH region. All four regions including Beijing-Tianjin-Hebei, Inner Mongolia, Xinjiang, and Qinghai show a significant upward trend in both solar radiation and PV generation potential from 2020 to 2060.

Due to computational limitations, we only simulated solar radiation and its key determinants in four representative months per year under the carbon-neutral emissions scenario, and used the simulation results from these four months to estimate the annual averaged solar radiation and total PV power generation. In addition, the PV development area in this model is based on static land-use data and does not account for potential reductions caused by urban expansion and policy changes between 2015 and 2060. Similarly, it does not consider increases of PV power potential from distributed applications such as building-integrated PV or rooftop installations. Therefore, the results of this study have a certain uncertainty. Future research will extend simulations to full year under more emission scenarios and further improve the PV estimation model by incorporating future PV industry developments and policy evolution. We will develop more comprehensive and accurate high-resolution solar resource datasets tailored to diverse emission pathways.

## CRediT authorship contribution statement

**Daoming Zhu:** Writing - original draft, preparation, Methodology. **Xinghong Cheng:** Conceptualization, Methodology, Writing - review & editing. **Yanbo Shen:** Investigation, Validation, Supervision. **Chunsong Lu and Duanyang Liu:** Resources, Methodology. **Shuqi Yan and Naifu Shao:** Methodology. **Zhongfeng Xu:** Data curation. **Jida Peng and Bing Chen:** Software.

# Declaration of competing interest

The authors declare that they have no known competing financial interests or personal relationships that could have appeared to influence the work reported in this paper.

# Acknowledgments

This work was supported by the Beijing-Tianjin-Hebei Environmental Comprehensive Management National Science and Technology Major Project, administered by the Ministry of Ecology and Environment (grant no. 2025ZD1204905). The measured surface radiation, air temperature and cloud data and the DPEC emission inventory are available from the Chinese Meteorological Administration data portal (https://data.cma.cn), and Tsinghua University data portal (http://meicmodel.org.cn/?page_id=1917), respectively. The 6-hourly multi-model integrated CMIP6 dataset is from the research [22]. The aerosol optical depth data are available from https://ladsweb.modaps.eosdis.nasa.gov/missions-and-measurements/products/MCD19A2. The simulated data and other information that support the plots within this paper and other findings of this study are available from the corresponding author upon reasonable request. The authors acknowledge Tsinghua University for providing the emissions inventory datasets.